\documentclass[12pt, a4paper]{article}
\usepackage[english]{babel} 

\usepackage{latexsym,bm}
\usepackage{amsmath,amsfonts}
\usepackage{amssymb}
\usepackage{mathtools}

\usepackage{tensor}

\usepackage{hyperref}

\usepackage{subfigure}
\usepackage[subfigure]{tocloft}
\usepackage{float}
\usepackage{graphicx}

\usepackage{dcolumn}

\usepackage{geometry}
\usepackage[affil-it]{authblk}  
\usepackage{color}
\usepackage{sidecap}
\usepackage{tocloft}

\usepackage{indentfirst} 

\usepackage{geometry}

\usepackage[format=hang, font={small, it}, labelfont=bf]{caption}

\usepackage[square,numbers, sort&compress]{natbib}
\usepackage{xcolor}                                                   	%
\usepackage{amscd}							%
\usepackage{mathrsfs}						%
\usepackage{tikz-cd}							%
\usepackage[all,cmtip]{xy}						%
\usepackage{float}							%
\usepackage{mathtools}						%
\usepackage{dsfont}

\usepackage{booktabs}

\numberwithin{equation}{section}
\allowdisplaybreaks[4] 
\makeatletter
\renewcommand*{\@fnsymbol}[1]{\ifcase#1\or$\dagger$\else\@arabic{\numexpr#1-1\relax}\fi}
\makeatother

\usepackage[mathlines]{lineno}

\begin{document}

\title{Kinetic Equation for Stochastic Vector Bundles}

\author[1,2,$\dagger$]{De-yu ZHONG}
\author[1]{Guang-Qian WANG}
\affil[1]{State Key Laboratory of Hydroscience and Engineering, Tsinghua University, Beijing, 100084, China}
\affil[2]{Joint-Sponsored State Key Laboratory of Plateau Ecology and Agriculture, School of Water Resources and Electric Power, Qinghai University, Xining, 810016, China}

\affil[$\dagger$]{Corresponding author: zhongdy@tsinghua.edu.cn}




\date{}


\maketitle

\begin{abstract}
The kinetic equation is crucial for understanding the statistical properties of stochastic processes, yet current equations, such as the classical Fokker-Planck, are limited to local analysis. This paper derives a new kinetic equation for stochastic systems on vector bundles, addressing global scale randomness. The kinetic equation was derived by cumulant expansion of the ensemble-averaged local probability density function, which is a functional of state transition trajectories.  The kinetic equation is the geodesic equation for the probability space. It captures global and historical influences, accounts for non-Markovianity, and can be reduced to the classical Fokker-Planck equation for Markovian processes. This paper also discusses relative issues concerning the kinetic equation, including non-Markovianity, Markov approximation, macroscopic conservation equations, gauge transformation, and truncation of the infinite-order kinetic equation, as well as limitations that require further attention.
\end{abstract}


\tableofcontents

%
%
%
\section{Introduction}\label{sec-1}
Stochastic processes are prevalent in natural phenomena and human societies, characterised by the inherent randomness in all or part of the variables involved. As a result, our understanding of these processes is primarily based on their statistical properties rather than deterministic outcomes \cite{Bhattacharya2021}. The analysis of such processes necessitates the a priori determination of the probability density function (PDF), or the distribution function in classical statistical mechanics. The equations that govern the PDF's evolution are known as kinetic equations in statistical mechanics \citep{balescu, Reeks1991Kinetic}. Among the most recognised kinetic equations are the Boltzmann equation and the Fokker-Planck equation (FPE), along with their numerous adaptations, which find application across a diverse range of scenarios in physics, chemistry, mechanics, and even the social sciences \citep{risken1984fokker, van1992stochastic}.

The Boltzmann equation was initially derived from studying behaviours of dilute gas molecules by Ludwig Boltzmann; detailed reviews can be found in \citep{balescu}. It describes the time evolution of the PDF for a collection of molecules in velocity phase space, which is characterised by the position and velocity vectors of the molecules. The dynamics of the PDF are governed by the external forces applied to the particles and the internal collisions occurring among them. Although the Boltzmann equation was primarily developed with molecular motion, its principles have been adapted to study other systems where particle collisions play a significant role. For instance, it has been successfully applied to analyse granular or debris flows, where the collisions among solid particles are the dominant mechanism \citep{rapid-granular-flows}. Despite its origins in the context of gases, the Boltzmann equation's conceptual framework has proven to be a powerful tool for understanding the statistical behaviour of various particle systems.

The FPE is another crucial advance in statistical mechanics after the Boltzmann equation. The FPE is not only a milestone in studies of non-equilibrium statistical mechanics \cite{balescu}, but also a significant foundation in studies of general stochastic processes \citep{van1992stochastic}. It shows that the time rate of change of PDF arises from two different effects: drift caused by convection and diffusion caused by random forcing. The FPE carries much more information about the random forcing of a stochastic system than the Boltzmann equation, so far as the diffusion of system states is considered. However, the classical FPE suffers difficulties in dealing with non-Markovian processes, which are common in both natural phenomena and industrial or social procedures \citep{zwanzig2001nonequilibrium, DEDECKER2020124269, DEPPMAN2023137752}. Modifications of the classical FPE are abundant in the literature. For instance, the work presented in \citep{zwanzig2001nonequilibrium} proposed the projection approach, resulting in an integral-differential equation for PDF capable of considering historical effects. Other studies reported in \citep{PhysRevE.61.1247, KHAN2005183, ZABROCKI2006349, OLLA200651, BOLIVAR20113095} are also typical to account for memory effects. Since the FPE also plays a central part in studying turbulence and turbulent particle-laden flows, improvements on the FPE were also contributed by scientists in fluid mechanics and multiphase dynamics \citep{Reeks1991Kinetic, reeks1992continuume, Hyland1999PDF, zaichik2004probability, reeks_2005}. For example, the Lagrangian history direct interaction (LHDI) approach was proposed in \citep{reeks1992continuume} to account for stochastic and historical effects on particle motion; differently, the functional method was adopted in \citep{Hyland1999PDF} by which the long-time correlations can be taken into account.  

The FPE has been derived through various methodologies, among which the Kramers-Moyal expansion of the master equation, truncated at the second-order moments for Markovian processes, is a notable example. This approach is popular and cited in several important works, including \citep{risken1984fokker, van1992stochastic, Sto-m-cgispinGardiner}. In contrast, we introduced a new method for deriving a kinetic equation that applies to non-Gaussian, non-stationary, and non-Markovian processes, as detailed in \citep{ZHONG-2022-KINETICEQAUTION}. The local PDF was redefined as a functional of state transition trajectories within velocity phase space, different from the traditional treatment where the local PDF was represented solely as a function of states, as seen in \citep{risken1984fokker, Reeks1991Kinetic, reeks1992continuume, Hyland1999PDF, zaichik2004probability, reeks_2005, SBPOPE-TURBULENTFLOWS}. This formulation elucidates that the coefficients of the kinetic equation are expressed in terms of cumulants related to state transition paths, rather than the jump moments of system states in the Kramers-Moyal expansion. This distinction is pivotal when accounting for the historical dependence that influences the statistical properties of non-Markovian processes. Applying this kinetic equation to the diffusion of particles in turbulent flows has successfully captured complex statistical properties, such as Reynolds stresses of solid particles in turbulent flows. For the details of this approach, reference is made to \citep{ZHONG-2022-KINETICEQAUTION}.

However, the kinetic equation presented in \citep{ZHONG-2022-KINETICEQAUTION} assumes the entire phase space to be a vector space, specifically an $n$-dimensional Euclidean space. This approach is limiting because the configuration space is described better as a general manifold in many cases within both theoretical and applied research, as noted in \citep{Arnold1989, Shapere_Wilczek_1989}. For configuration space as a general manifold, the velocity (or momentum) phase space of it is the tangent (or cotangent) space of the configuration space, which only locally resembles $\mathbb{R}^n$ but not globally. A global stochastic analysis is essential to comprehend more complex stochastic systems, as highlighted in \citep{Gliklikh2011}. Consequently, developing kinetic equations valid for general manifolds has become a significant research priority.

This paper aims to derive a kinetic equation that applies to stochastic systems on general vector bundles. We begin with a review of kinetic equations in the context of Euclidean space in Section \ref{sec-2}, encompassing both the Kramers-Moyal expansion for Markov processes and the cumulant expansion for non-Markovian processes, as presented in \citep{ZHONG-2022-KINETICEQAUTION}. In Section \ref{sec-3}, we describe the mathematical framework of stochastic vector bundles, which serves as the groundwork for deriving our kinetic equation. Subsequent sections, \ref{sec-4} and \ref{sec-5}, delve into key aspects of the kinetic equation, addressing the geometric characteristics of stochastic vector bundles, the implications of non-Markovianity, and the Markov approximation, among other topics. The paper concludes with a comprehensive summary in Section \ref{sec-6}, encapsulating the main findings, contributions and limitations that require further investigation.

\section{Brief Review of Kinetic Equation for \texorpdfstring{$\mathbb{R}^n$}{}}\label{sec-2}
\subsection{Moment Expansion of Kramers and Moyal}\label{sec-2-1}
Previous studies have used several approaches to derive the kinetic equation for stochastic processes. Some of the commonly cited ones include those presented in \citep{risken1984fokker}, \cite{van1992stochastic}, and \citep{zwanzig2001nonequilibrium}. One of the popular methods is the local PDF (or fine-grained PDF) approach \citep{Hyland1999PDF, SBPOPE-TURBULENTFLOWS, ZHONG-2022-KINETICEQAUTION}. Typically, for a stochastic process $Y(t)\in \mathbb{R}^n$, the local PDF is defined as a functional $\chi: \mathbb{R}^n\times \mathbb{R}^n \to \mathbb{R}$. It is
\begin{equation}\label{eq-2-1}
\chi(Y, y)=\chi(|Y(t)-y|)
=
\begin{cases}
\infty, &\ Y(t)=y;\\
0, &otherwise.
\end{cases}
\end{equation}
In Eq.~\eqref{eq-2-1}, $|a-b|$ represents the Euclidean distance between two points $a$ and $b$, which belong to the $n$-dimensional Euclidean space, $\mathbb{R}^n$. The local PDF $\chi(y, Y)$ is responsible for determining the likelihood that a stochastic process $Y(t)$ attains a particular point $y \in \mathbb{R}^n$ at some specific time $t$. Although the Dirac delta function is usually employed as the local PDF $\chi$ \citep{Hyland1999PDF, SBPOPE-TURBULENTFLOWS}, there are alternative formulations available \citep{zhang1994averaged}. Consequently, we adopt a generic functional form for $\chi$ in Eq.~\eqref{eq-2-1} to characterise the local PDF.

The local PDF $\chi$ is employed to determine the mesoscale (or coarse-grained) PDF, or $f(y, t)$, of $Y(t)$. This is achieved by implementing the ensemble average on $\chi$ \citep{SBPOPE-TURBULENTFLOWS, ZHONG-2022-KINETICEQAUTION}. The equation for determining $f(y,t)$ is as follows:
\begin{equation*}
f(y,t) =\frac{ \int_{V_y} \chi(|Y(t)-y|) \mathrm{d}Y}{ \int_{V_y} \mathrm{d}Y} 
\equiv \langle \chi(|Y(t)-y|) \rangle.
\end{equation*}
In this equation, the ensemble average is denoted by $\langle  \rangle$; $\mathrm{d}Y=\mathrm{d}Y^{1}\cdots\mathrm{d}Y^{m}$ represents the infinitesimal volume of $\mathbb{R}^n$ centered at $Y$, and $V_y\in \mathbb{R}^n$ is a volume that contains the point $y$.

To derive the Kramers-Moyal expansion for the PDF $f(y,t)$, one typically expands the local PDF $\chi(|Y(t) - y|)$ using a Taylor series and then performs ensemble averaging on the resulting expression. This approach assumes that the stochastic processes under consideration are Gaussian and Markovian \citep{risken1984fokker}. The Kramers-Moyal expansion of $f(y,t)$ is: 
\begin{equation}\label{eq-2-2}
\frac{\partial f(y, t )}{\partial t}=L_{KM}f(y, t ),
\end{equation}
where
\begin{equation*}
\begin{split}
&L_{KM}
=\sum_{n=1}^{\infty}\frac{(-1)^n}{n!} D^{(n)}_{KM}, \\
&D^{(n)}_{KM}
=\lim_{\tau\to 0}\frac{\langle \Delta Y^{\mu_1}\cdots \Delta Y^{\mu_n}\rangle }{\tau} \frac{\partial^{n}}{\partial y^{\mu_1}\cdots \partial y^{\mu_n}}, \\
&\Delta Y
=Y(t)-Y(t-\tau),
\end{split}
\end{equation*}
with $\langle \Delta Y^{\mu_1}\cdots \Delta Y^{\mu_n}\rangle$ denoting the $n$-th order moments of Markovian jumps in the stochastic process $Y(t)$.

\subsection{Cumulant Expansion of Zhong et al.}\label{sec-2-2}

Considering historical effects on stochastic processes is a challenging problem \citep{zwanzig2001nonequilibrium, reeks1992continuume, ZHONG-2022-KINETICEQAUTION}. 
Zhong et al. \citep{ZHONG-2022-KINETICEQAUTION} proposed a new local PDF, of which the ensemble average approach is based on state transition paths of $\mathbb{R}^{n}$. It is 
\begin{equation}\label{eq-2-3}
\chi(Y, y)=\chi(|Y(t|y_0, 0)-y|)
=
\begin{cases}
\infty, &Y(t|y_0, 0)=y,\\
0, &otherwise,
\end{cases}
\end{equation} 
of which $Y(t|y_0, 0)$ represents the paths of a system to change its state from $y_0$ at time $0$ to $Y$ at time $t$. The local PDF conserves along the trajectory $Y(t|y_0, 0)$ \citep{ZHONG-2022-KINETICEQAUTION}; it is to say that
\begin{equation}\label{eq-2-4}
\frac{\partial \chi}{\partial t}+\dot{Y}\cdot\frac{\partial \chi}{\partial y}=0.
\end{equation}
Different from Eq.~\eqref{eq-2-1}, the ensemble average of $\chi$ over state-transition-paths are conditional PDF \citep{ZHONG-2022-KINETICEQAUTION}, i.e., 
\begin{equation}\label{eq-2-5}
f(y, t|y_0, 0) \equiv \langle  \chi(|Y(t|y_0, 0)- y|) \rangle.
\end{equation}

Using the local PDF given by Eq.~\eqref{eq-2-4},  they derived a kinetic equation for stochastic systems of $\mathbb{R}^n$  \cite{ZHONG-2022-KINETICEQAUTION}:
\begin{equation}\label{eq-2-6}
\frac{\partial f(y, t|y_0, 0)}{\partial t}=\sum_{n=1}^{\infty}\frac{(-1)^{n}}{n!}\nabla^{n}D^{(n)}f(y, t|y_0, 0),
\end{equation}
of which 
\begin{equation*}
D^{(n)}=\frac{\partial}{\partial t}
\frac{
\langle\langle S^{n}\rangle\rangle
}{n!}, 
\ S=\int_{0}^{t} \dot{Y}(\tau)\mathrm{d}\tau,
\end{equation*}
and $\langle\langle S^n\rangle\rangle\equiv \langle\langle \overbrace{S\otimes \cdots \otimes S}^{n-tuples}\rangle\rangle$ represents the $n$-th order cumulant regarding $S$.   
The PDF $f(y, t)$ can be derived similarly. Details of the formulation for Eq.~\eqref{eq-2-6} and $f(y, t)$ are referred to \cite{ZHONG-2022-KINETICEQAUTION}, of which discussions and applications of Eq.~\eqref{eq-2-6} to particle-laden turbulent flows are also presented.

In Eq.~\eqref{eq-2-2}, the jumps are quantified by $ \Delta Y= Y(t)-Y(t-\tau)$. In the context of the Fokker-Planck equation, it is necessary for $\tau\to 0$, implying that the jumps must be infinitesimally small to ensure the jump moments
\begin{equation*}
M^{(n)}=\lim_{\tau\to 0}\frac{\langle\Delta Y^{\mu_1}\cdots Y^{\mu_n}\rangle}{\tau}
\end{equation*}
are meaningful. In contrast, Eq.~\eqref{eq-2-6} introduces $S=\int_0^t \dot{Y}\mathrm{d}\tau$ to capture the cumulative distance of state transitions from $y_0$ to $Y(t|y_0,0)$. The distinction between $\Delta Y$ and $S$ is clear: $\Delta Y= Y(t)-Y(t-\tau)$ with $\tau\to 0$ represents local infinitesimal displacements, whereas $S=\int_0^t \dot{Y}(\tau|y_0,0)\mathrm{d}\tau$ encapsulates the entire history of transitions along the path $Y(\tau|y_0,0)$ for $0\le\tau\le t$. At the same time, for small correlation time scales where $T_L=\tau\to 0$, $S=\int_0^t \dot{Y}(\tau|y_0,0)\mathrm{d}\tau$ approximates $Y(t)-Y(t-\tau)$, leading to the equivalence of $S$ and $\Delta Y$ in the context of Markovian processes.

\section{Kinetic Equation for Stochastic Vector Bundles}\label{sec-3}
\subsection{Stochastic Vector Bundles}\label{sec-3-1}

Let $(\Omega, \mathcal{F}, \mathbb{P})$ denote a  probability space with sample space $\Omega$, $\sigma$-algebra $\mathcal{F}$, and probability measure $\mathbb{P}$; let $E=F\times_GV$ denote a vector bundle with the base manifold $M$ of dimension $n$ and typical fiber $V\cong \mathbb{R}^n$ associated with a frame bundle $F$ with structure group $G$. The probability space $(\Omega, \mathcal{F}, \mathbb{P})$ is related to the vector bundle $E$ by the following mappings:

\begin{enumerate}

\item The probability space $(\Omega, \mathcal{F})$ is related to the smooth manifold $M$ by the map $\gamma$ defined to be a stochastic process $\gamma: (\Omega, \mathcal{F})\hookrightarrow M$, which embeds $(\Omega, \mathcal{F})$ into the manifold $M$. 

\item There is a surjective projection $\pi: E \to M$ and an injective map $Y: M \to E: p\mapsto Y(p)$, known as the section of $E$, satisfying $\pi\circ Y=\mathds{1}_{M}$ where $\mathds{1}_{M}$ is an identity map.

\end{enumerate}
Strictly speaking, $\gamma: (\Omega, \mathcal{F})\hookrightarrow M$ should be understood as $\gamma: \{\mathcal{F}\}_{t\in T}\hookrightarrow M$, where  $\{\mathcal{F}\}_{t\in T}\subseteq \mathcal{F}$ is the filtration of $\mathcal{F}$ adapted to $\gamma$. Because we are not focused on the structures of $(\Omega, \mathcal{F})$, we will not distinguish  $\{\mathcal{F}\}_{t\in T}\subseteq \mathcal{F}$ from $\mathcal{F}$ herein and after.

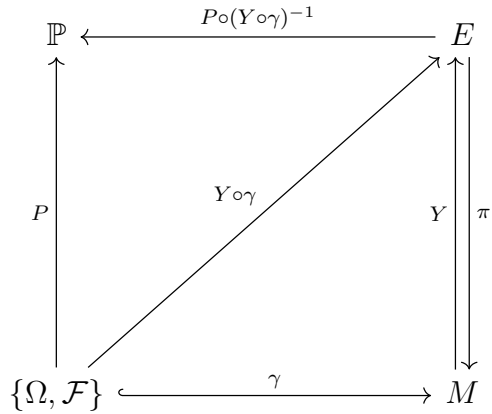
\begin{figure}[h]
\centering

\begin{tikzcd}[ column sep=10em, row sep=10em]
 \mathbb{P}  & E 
	\arrow[xshift=0.5ex]{d}{\pi} 
	\arrow[swap]{l}{P\circ(Y\circ\gamma)^{-1}}\\
\{\Omega, \mathcal{F}\} 
	\arrow{u}{P}
	\arrow[hook]{r}{\gamma}
	\arrow{ur}{Y\circ \gamma}
& M \arrow[xshift=-0.5ex]{u}{Y}
\end{tikzcd}
\caption{\label{fig: Fig1} The commutative diagram for stochastic vector bundles, where $E=F\times_GV$ is a vector bundle associated to a frame bundle $F$ with typical fiber $V\cong \mathbb{R}^n$; $M$ is the base manifold of $E$ with the canonical projection $\pi: E\to M$; $\{\Omega, \mathcal{F}, \mathbb{P}\}$ is a measurable probability space with sample space $\Omega$, $\sigma$-algebra $\mathcal{F}$, and probability measure space $\mathbb{P}$; $\gamma$ is an embedding of $\{\Omega, \mathcal{F}\}$ into $M$; $Y$ is a section of  $\pi: E\to M$, and $Y\circ \gamma: \{\Omega, \mathcal{F}\}\to E $ defines a stochastic vector with the probability measure $P\circ (Y\circ\gamma)^{-1}: E\to \mathbb{P}$ on $E$.}
 \end{figure}

With $\gamma$ and $Y$, we can define a composition map 
\begin{equation*}
Y\circ \gamma: (\Omega, \mathcal{F})\to E,
\end{equation*}
which maps the probability measure space $\mathbb{P}$ to the vector bundle $\pi: E\to M$, and moreover, let the probability $P\in \mathbb{P}$ work as a map that
\begin{equation*}
P\circ (Y\circ \gamma)^{-1}: E\to \mathbb{P},
\end{equation*}
then we constructed a stochastic vector bundle, as shown in Fig.~\ref{fig: Fig1}. 

\subsection{Probability Distribution Function}\label{sec-3-2}

Considering a stochastic vector bundle $\pi: E \to M$, where $\gamma: [0,t] \to M$ is a path with $\gamma(0) = p \in U_p \subset M$ and $\gamma(t) = q \in U_q \subset M$; $U_p$ and $U_q$ are two open neighborhoods around $p$ and $q$ ($p \neq q$), respectively. Given $y_0 \in \pi^{-1}(p) = E_p$ and $y \in \pi^{-1}(q) = E_q$, we have two distinct points in the fibres of the vector bundle $E$ over $p$ and $q$. Assuming that the vector bundle $E$ represents state space of a dynamical system, we can interpret a section $Y: M \to E$ as a collection of states of the system. If $Y$ is a stochastic process, then $Y(\gamma(t))$ can be viewed as a stochastic trajectory or state transition path in the bundle. This path starts at the state represented by $y_0$ when $t = 0$ and evolves to the state represented by $Y(\gamma(t)) = y$ as $t$ transitions from $0$ to $t$, following the path $\gamma$ in the base manifold $M$. It is 
\begin{equation*}
Y(\gamma(t))=Y(t|y_0, 0),
\end{equation*}
where $Y(t|y_0, 0)$ means that the system is at the state $y_0$ at time $0$ while a the state $Y$ at time $t$.

One might propose using a local PDF $\chi$, akin to Eq.~\eqref{eq-2-3}, to determine $P$ for $Y(t|y_0, 0)$ within the vector bundle. However, Eq.~\eqref{eq-2-3} is not well-defined for general vector bundles. The issue arises because the set of random trajectories $\{Y(t|y_0, 0) \,|\, \forall Y(0)=y_0 \in E\}$, for each realization $\omega \in \Omega$, does not guarantee that both $Y$ and $y$ lie within the same open neighborhood in the vector space $E_q$. This discrepancy renders it impossible to quantify the distance between stochastic points $\{Y\}$ and $y$ using Eq.~\eqref{eq-2-3}. 

This problem was resolved by introducing the concept of parallel transport developed in differential geometry \citep{Nakahara2003GeometryTA}. Let $G$ to be a Lie group, and set $P_{\gamma}^t$ to be a map that:
\begin{equation*}
G\ni P_{\gamma}^t: E\to E,
\end{equation*}
by which $Y(t|y_0, 0)$  is transported back to $E_{p}$ along $\gamma(t)$ in a parallel manner
\begin{equation*}
Z(t|y_0, 0)= P_{\gamma}^{-t} Y(t|y_0, 0).
\end{equation*}

Furthermore, define $\rho: E\times E\to \mathbb{R}$ as a distance functional measuring the distance between two points in an open subset $E_{p}$. Without loss of generality, it is assumed that the distance from $Z$ to any given point $z\in E_{p}$ is 
\begin{equation*}
 \rho(Z(t), z)=|Z(t)-z|=\prod_{k=1}^n |Z^{k}(t)-z^{k}|.
\end{equation*}
Obviously, for $y$, $z$, and $Z\in E_p$, it possesses the properties of :  
\begin{enumerate}
\item $\rho(Z, z)=\rho(z, Z)$.
\item $\rho(Z, z)\ge 0$. 
\item $\rho(Z, z)\le \rho(Z, y)+\rho(y, z)$. 
\end{enumerate}
These properties of $\rho$ prove it is indeed a distance functional. 

With the parallel transport $P_{\gamma}^t$ and the distance functional $\rho$, the local PDF $\chi$ is defined as  
\begin{equation}\label{eq-3-1}
\chi=
\chi\circ \rho(Z, z)
=
\begin{cases}
\infty,  &\rho(Z, z)=0,\\
0, &otherwise,
\end{cases}
\end{equation}
and by which the local probability of $Z \in \mathrm{d}z$ is 
\begin{equation*}
P(Z\in \mathrm{d}z)=\langle\chi(Z, z)\rangle \mathrm{d}z.
\end{equation*}
$P(Z\in \mathrm{d}z)$ measures the local probability to find $Z$ in an infinitesimal volume $ \mathrm{d}z=\mathrm{d}z^{\mu_1}\wedge\cdots\wedge\mathrm{d}z^{\mu_n}$. It is found that  
\begin{equation}\label{eq-3-2}
\begin{split}
P(Z(t|y_0,0)\le z)
&=P(P_{\gamma}^{-t} Y(t|y_0,0)\le z) \\ 
&= P(Y(t|y_0,0)\le P_{\gamma}^t z) \\ 
&= P(Y(t|y_0,0)\le y),
\end{split}
\end{equation}
which indicates that parallel transport does not change the probability measure $P$. It should be emphasized that $P(Z\in \mathrm{d}z)$ is understood as $P\circ (Z\circ\gamma)^{-1}: E_z \subseteq E\to \mathbb{P}$ in order to simplify equations.

\subsection{Conservation Equation for Local PDF}\label{sec-3-3}

Following the same procedure as reported in \citep{ZHONG-2022-KINETICEQAUTION}, for a divergent-free system, i.e., 
\begin{equation*}
\frac{\partial \dot{Z}^{\mu}}{\partial z^{\mu}}=0.
\end{equation*}
The conservation equation of the local PDF $\chi$ in the vector space is given by 
\begin{equation*}
\frac{\mathrm{d}\chi\circ \rho(Z(t), z)}{\mathrm{d}t}
=
\frac{\partial \chi\circ \rho(Z(t), z)}{\partial t}+D\chi(\rho)\cdot D_1\rho(Z(t), z)\cdot \dot{Z}(t)=0,
\end{equation*}
where $D$ stands for the derivative, whereas the subscripts $j$ ($j=1, 2$) of $D$ stand for the partial derivatives with respect to the $j$-th variable in multi-dimensional functions.

It is easy to verify that the derivatives of $\chi\circ\rho(Z, z)$  with respect to $Z$ and $z$ are antisymmetric, which means that 
\begin{equation*}
D\chi(\rho)\cdot D_1\rho(Z, z)=-D\chi(\rho) \cdot D_2\rho(Z, z).
\end{equation*}
Moreover, using the chain role for derivatives of composition functions, it is found that 
\begin{equation*}
\begin{split}
\dot{Z}(t)\cdot D\chi(\rho)\cdot D_1\rho(Z, z)
&=-\dot{Z}(t)\cdot D\chi(\rho) \cdot D_2\rho(Z, z)\\
&=-(\dot{Z}(t)\cdot D_2) (\chi\circ\rho)(Z, z).
\end{split}
\end{equation*}

As techniques developed in differential geometry will significantly reduce the challenges in global analysis, theories and approaches of differential manifolds will be employed in the following formulation. The operator $\dot{Z}(t)\cdot D_2$ acting on $\chi\circ\rho$  in the above equation is, in essence, a directional derivative. In the terminology of differential geometry \citep{Nakahara2003GeometryTA}, it is expressed as 
\begin{equation*}
\dot{Z}\cdot D_2
=\dot{Z}(z^{\mu})\frac{\partial}{\partial z^{\mu}}
\equiv
\nabla Z (\dot{\gamma}(t)),
\end{equation*}
where $\{\dot{Z}(z^\mu)\}$ is the component of the vector $\dot{Z}$ under the basis $\{\partial/\partial z^{\mu}\}$; $\nabla: \Gamma(M\times E)\to \Gamma(E)$ is a connection 1-form given as follows. 

The vector $Z$ is understood as the parallel transport of $Y$ along the curve $\gamma$, expressed as $Z=P_{\gamma}^{-t}Y$. Simultaneously, the covariant derivative of $Y$ is given by $\nabla Y = -(\mathrm{d}+A_{\gamma}) Y$, where $A_{\gamma}$ denotes the local connection 1-form. In the context of flat space, $A_{\gamma}$ vanishes, simplifying the covariant derivative to $\nabla = -\mathrm{d}$. Applying $\nabla$ on $Z$, one reads that  
\begin{equation*}
\begin{split}
\nabla Z&=
\mathrm{d}P_{\gamma}^{-t}Y+P_{\gamma}^{-t}\nabla Y\\
&=-[P_{\gamma}^{-t}\mathrm{d}P_{\gamma}^{t}P_{\gamma}^{-t}Y+P_{\gamma}^{-t}(\mathrm{d}+A_{\gamma})Y]\\
&=-[P_{\gamma}^{-t}\mathrm{d}P_{\gamma}^{t}Z+P_{\gamma}^{-t}(\mathrm{d}+A_{\gamma})P_{\gamma}^{t}Z]\\
&=-[\mathrm{d}+\underbrace{(P_{\gamma}^{-t}\mathrm{d}P_{\gamma}^{t}+\mathrm{Ad}_{P_{\gamma}^{-t}} A_{\gamma})}_{=\omega_{\gamma}}]Z\\
&\equiv -\omega Z,
\end{split}
\end{equation*}
where
\begin{equation*}
\omega=\mathrm{d}+\omega_{\gamma},
\end{equation*}
with the local connection $\omega_{\gamma}$ defined as 
\begin{equation*}
\omega_{\gamma}=P_{\gamma}^{-t}\mathrm{d}P_{\gamma}^{t}+\mathrm{Ad}_{P_{\gamma}^{-t}} A_{\gamma},
\end{equation*}
usually termed as the gauge transformation of $A_{\gamma}$ \citep{Nakahara2003GeometryTA}. With the help of the gauge transformation,  we found that 
$
\nabla=-\omega,
$
which shows that $\nabla$ is a connection 1-form as expected. 

Finally, the directional derivative acting on $\chi$ is expressed concisely with the aid of the connection 1-form $\omega$ as follows:
\begin{equation*}
\dot{Z}\cdot D_2\chi=\nabla Z(\dot{\gamma}(t))\chi=-Z\otimes \omega(\dot{\gamma}(t))\chi.
\end{equation*}
It should be noted that $\nabla Y$ is typically defined as $\nabla Y = (\mathrm{d}+A_{\gamma}) Y$ in differential geometry. However, a negative sign is usually included before the parentheses for convenience in physics and fluid dynamics. 

The parallel transport operator $P_{\gamma}^t$  is determined by the requirement that, if $Z$ is a parallel section, then the following identity holds \citep{Nakahara2003GeometryTA}:
\begin{equation*}
Z^{*}\omega_{\gamma}(\dot{\gamma}(t))=
\omega_{\gamma}(Z_{*}\dot{\gamma}(t))=
0, 
\end{equation*}
and it leads to
\begin{equation*}
\frac{\mathrm{d}P_{\gamma}^t}{\mathrm{d}t}=-\omega_{\gamma}(Z_{*}\dot{\gamma}(t)) P_{\gamma}^t.
\end{equation*} 
Since that $\omega_{\gamma}(Z_{*}\dot{\gamma}(t))=Z^{*}\omega_{\gamma}(\dot{\gamma}(t))=A_{\gamma}(\dot{\gamma}(t))$  \citep{Nakahara2003GeometryTA}, the parallel translation operator $P_{\gamma}^t$ is obtained from the above differential equation iteratively as
\begin{equation}\label{eq-3-3}
P_{\gamma}^t=\overleftarrow{T} 
\mathrm{e}^{
			-\int_0^{t} \mathrm{d}\tau A_{\gamma}(\tau)},
\end{equation}
where $\overleftarrow{T}$ is the time ordering operator \citep{Nakahara2003GeometryTA}.

Using the connection 1-form $\omega$, we defined a vector-valued 1-form $\mathcal{L}=\mathrm{Hom}_{\mathbb{R}}(TM, TE)$:
\begin{equation*}
\mathcal{L} 
\equiv Z\otimes \omega, 
\end{equation*}
which is usually referred to as the Liouville operator in statistical mechanics \citep{zwanzig2001nonequilibrium}, and therefore, the variation of the local PDF is given in a simple operator form as:
\begin{equation}\label{eq-3-4}
\begin{split}
\frac{\partial \chi  }{\partial t}
+\mathcal{L}(\dot{\gamma}(t)) \chi=0.  
\end{split}
\end{equation}

Eq.~\eqref{eq-3-4} shares a similar form with equations derived in \citep{Hyland1999PDF, zwanzig2001nonequilibrium, ZHONG-2022-KINETICEQAUTION}, yet it possesses a crucial distinction: it is coordinate-independent. This means that the equation maintains its form regardless of coordinate transformations, ensuring the conservation of the local PDF across different coordinate systems. This property is essential as it allows for applying the local PDF to general manifolds, where the choice of coordinates can be arbitrary and may change from point to point. The coordinate-independence of Eq.~\eqref{eq-3-4} thus provides a foundation for analysing systems on curved spaces and complex manifolds.


%
%
%
\subsection{Formulation of the Kinetic Equation}\label{sec-3-4}

Solving in an iterative manner \citep{risken1984fokker, ZHONG-2022-KINETICEQAUTION}, we found a formal solution to Eq.~\eqref{eq-3-4}. It is
\begin{equation}\label{eq-3-5}
\chi(Z, z)=G(\mathcal{L})\chi(y_0, z),
\end{equation}
where $G(\mathcal{L})$ is the time evolution operator, given by
\begin{equation*}
G(\mathcal{L})=\overleftarrow{T}\mathrm{exp}\left(-\int_0^t \mathcal{L}(\dot\gamma(\tau))\mathrm{d}\tau\right).
\end{equation*}
Before proceeding further, observing how Eq.~\eqref{eq-3-5} varies under diffeomorphisms is advisable.

Define an isomorphism in the vector space so that: 
\begin{equation*}
\psi: E_{p} \to E_{q}: z\mapsto y=\psi(z), 
\end{equation*}
which at the same time leads to the pull-back  
\begin{equation*}
\psi^{*}: T^{*}_qE\to T^{*}_pE
\end{equation*}
and the push-forward
\begin{equation*}
\psi_{*}: T_pE\to T_qE.
\end{equation*}
Let $U_z \subseteq E_p$ and $U_y \subseteq E_q$ be two open neighborhoods of $z\in E_p$ and $y\in E_q$, respectively, and $\psi(U_z)=U_y$. With the pull-back $\psi^{*}$, the integration of $\chi(Y, y)$ within $U_y$ can be transformed to that within $U_z$ \citep{Lee2012}. It is
\begin{equation*}
\begin{split}
\int_{U_y} \chi 
&=\int_{\psi(U_z)} \chi  
=\int_{U_z} \psi^{*} \chi. 
\end{split}
\end{equation*}

On the one hand, applying the ensemble average on both sides of the integration transformation, one finds that
\begin{equation}\label{eq-3-6}
\begin{split}
\int_{U_y} \langle\chi \rangle
&=\int_{U_z}\langle \psi^{*} \chi\rangle.
\end{split}
\end{equation}
On the otherhand, applying $\psi^{*}$ on both sides of Eq.~\eqref{eq-3-5}, one reads that 
\begin{equation}\label{eq-3-7}
(\psi^{*}\chi)(Z, z)=(\psi^{*}G)(\psi^{*}\chi)(y_0, z).
\end{equation}
At the same time, bearing in mind that ${G}(\mathcal{L})$ is a function of a vector-valued 1-form $\mathcal{L}$, so that \citep{Nakahara2003GeometryTA}
\begin{equation*}
(\psi^{*}{G})(\mathcal{L})={G}(\psi_{*}\mathcal{L}).
\end{equation*}
Finally, integrating Eq.~\eqref{eq-3-7} with respect to $z$ in the subspace $U_z \subseteq E_p$, applying ensemble average, and using the relationship Eq.~\eqref{eq-3-6}, one arrives at
\begin{equation}\label{eq-3-8}
\begin{split}
P(Y\in U_{y(t|y_0, 0)})
=\mathcal{G}(\psi_{*}\mathcal{L})P(Y\in U_{y(0|y_0, 0)}),
\end{split}
\end{equation}
of which the probability is   
\begin{equation*}
P(Y\in U_{y(t|y_0, 0)})
\equiv \int_{U_y} \langle \chi \rangle(Y, y) \mathrm{d}y, 
\end{equation*}
and 
\begin{equation*}
\begin{split}
&\mathcal{G}(\mathcal{L}(\dot{\gamma}(t)))
=\langle{G}(\mathcal{L}(\dot{\gamma}(t))) \rangle\\
&\equiv \overleftarrow{T}
\mathrm{exp}
\left(
\sum_{n=1}^{\infty}\frac{(-1)^n}{n!}
\left\langle 
\left\langle
\left(
\int_0^t \mathcal{L}(\dot\gamma(\tau)) \mathrm{d}\tau
\right)^n
\right\rangle
\right\rangle
\right).
\end{split}
\end{equation*}
with $\langle\langle\rangle\rangle$ represents the cumulant of stochastic variables \citep{ZHONG-2022-KINETICEQAUTION}. 
In addition, Eq.~\eqref{eq-3-7} also leads to 
\begin{equation}\label{eq-3-9}
\begin{split}
f(y, t|y_0, 0)
=\mathcal{G}(\psi_{*}\mathcal{L})f(y,0|y_0,0),
\end{split}
\end{equation}
of which $f(y,t|y_0, 0)$ denotes the conditional PDF. It shows that Eqs.~\eqref{eq-3-8} and  \eqref{eq-3-9} are topological invariant under diffeomorphism, which is significant for us to make the following transformation.  

As a special case of the diffeomorphism $\psi$, the parallel transport $P_{\gamma}^{t}$ leads to $y=\psi(z)=P_{\gamma}^{t}z$, with the push-forward $\psi_{*}$ determined as 
\begin{equation*}
\psi_{*}=\left(\frac{\partial y}{\partial z}\right)=P_{\gamma}^t.
\end{equation*}
Remembering that $\mathcal{L}(\dot{\gamma}(t))$ is a vector-valued 1-form, its  expression in the local coordinates of $T_pE$ is given by  
\begin{equation*}
\mathcal{L}(\dot{\gamma}(t))=\mathcal{L}[z^{\mu}]\frac{\partial}{\partial z^{\mu}},
\end{equation*}
and using the equivariant relation for vectors \citep{Nakahara2003GeometryTA}, the push-forward of $\mathcal{L}$ by $\psi_{*}=P_{\gamma}^t$ is
\begin{equation*}
\begin{split}
\psi_{*}\mathcal{L}
&=\left[\left(\mathcal{L}[z^{\mu}], \frac{\partial}{\partial z^{\mu}}\right)\right]P_{\gamma}^{t}
=
\left[\left(\mathcal{L}[z^{\mu}]P_{\gamma}^{t}, P_{\gamma}^{-t}\frac{\partial}{\partial z^{\mu}}\right)\right]\\
&=
\left[\left(\mathcal{L}[y^{\mu}], \frac{\partial}{\partial y^{\mu}}\right)\right]
=\mathcal{L}[y^{\mu}]\frac{\partial}{\partial y^{\mu}},
\end{split}
\end{equation*}
which shows that the right action of $P_{\gamma}^{t}$ as a Lie group moves $\mathcal{L}(\dot{\gamma}(t))\in T_pE$ to $T_qE$. 

Expanding Eq.~\eqref{eq-3-8} or \eqref{eq-3-9} as a Taylor series, we obtained that 
\begin{subequations}\label{eq-3-10} 
\begin{equation}\label{eq-3-10-a}
\frac{\partial P}{\partial t}=\mathscr{L}P,
\end{equation}
\text or
\begin{equation}\label{eq-3-10-b}
\frac{\partial f(y,t|y_0,0)}{\partial t}
=
 \mathscr{L}  f(y,t|y_0,0).
\end{equation}
\end{subequations} 
In Eqs.~\eqref{eq-3-10}, $ \mathscr{L} $ is the ensemble-averaged operator $\mathcal{L}$, given by
\begin{equation*}
\mathscr{L}
=\sum_{n=1}^{\infty}\frac{(-1)^n}{n!}\frac{\partial}{\partial t} \overleftarrow{T}
\left\langle 
\left\langle
\left(
\int \psi_{*} \mathcal{L}
\right)^n
\right\rangle
\right\rangle=\sum_{n=1}^{\infty}\frac{(-1)^n}{n!} D^{(n)},
\end{equation*}
of which 
\begin{equation*}
\begin{split}
 D^{(n)}
&\equiv
\frac{\partial}{\partial t} 
\frac{ \langle 
\langle
\overbrace{
{S} \cdots {S} 
}^{n-\text{tuples}}
\rangle
\rangle}{n!}
\cdot
\underbrace{\partial_y \otimes \cdots \otimes\partial_y}_{n-\text{tuples}}
=
\left(\frac{\partial}{\partial t} 
\frac{ \langle 
\langle
{S}^{\mu_1} \cdots {S^{\mu_n}} 
\rangle
\rangle}{n!}\right)
\frac{\partial}{\partial y^{\mu_1}} \cdots \frac{\partial}{\partial y^{\mu_n}}, \\
S
&\equiv\int_0^t \psi_{*}\mathcal{L}(\dot{\gamma}(\tau))\mathrm{d}\tau=S^{\mu}\frac{\partial}{\partial y^{\mu}},
\quad S^{\mu}
=
\int_0^t \mathcal{L}[y^{\mu}]\mathrm{d}\tau=\int_0^t \mathcal{L}^{\mu}\mathrm{d}\tau.
\end{split}
\end{equation*}
In the above derivation, we used the following identity \cite{ZHONG-2022-KINETICEQAUTION}:
\begin{equation*}
\begin{split}
\overleftarrow{T}
\left\langle\left\langle
\left(
\int_0^t \mathrm{d}\tau \mathcal{L}
\right)^n
\right\rangle\right\rangle
=
\frac{1}{n!}
\left\langle\left\langle
\left(
\int_0^t \mathrm{d}\tau \mathcal{L}
\right)^n
\right\rangle\right\rangle.
\end{split}
\end{equation*}

Eqs.~\eqref{eq-3-10-a} and \eqref{eq-3-10-b} are the kinetic equations derived for stochastic vector bundles. Although they strongly resemble the equation presented in \citep{ZHONG-2022-KINETICEQAUTION}, there is a fundamental distinction. These new equations apply to general vector bundles $E=F\times_GV$ instead of being limited to the vector space of $\mathbb{R}^n$. This extension to general vector bundles significantly broadens the scope of their applicability, allowing for the analysis of more complex geometric structures and dynamical systems.

\subsection{Component Representation of \texorpdfstring{$\mathscr{L}$}{}}\label{sec-3-5}

$\mathscr{L}$ is a function of the coefficient ${D}^{(n)}$, which can be interpreted geometrically as a tensor field and physically as a space derivative operator. The tensor field on the stochastic vector bundles is a function of the $n$th-order cumulant of $S$, given by
\begin{equation*}
\begin{split}
{D}^{(n)}
&=
\frac{1}{n!}\frac{\partial \langle\langle S^{\mu_1}\cdots S^{\mu_n} \rangle\rangle}{\partial t}
\frac{\partial}{\partial y^{\mu_1}}\otimes\cdots\otimes\frac{\partial}{\partial y^{\mu_n}}\\
&=
\int_0^{t}\cdots \int_{0}^{t}
\left\langle \left\langle
	\mathcal{L}^{\mu_1}(\dot{\gamma}(t))\cdots\mathcal{L}^{\mu_n}(\dot{\gamma}(\tau_n)) 
\right\rangle\right\rangle
\mathrm{d}\tau_{2}\cdots \mathrm{d}\tau_{n}
\frac{\partial}{\partial y^{\mu_1}}\otimes\cdots\otimes\frac{\partial}{\partial y^{\mu_n}}\\
&=
\int_0^t\cdots\int_0^t 
\left\langle \left\langle
\mathcal{L}^{\mu_1}\otimes\cdots\otimes\mathcal{L}^{\mu_n}
(\dot{\gamma}(t),\cdots, \dot{\gamma}(\tau_{n}))
\right\rangle\right\rangle
\mathrm{d}\tau_2\cdots\mathrm{d}\tau_n
\frac{\partial}{\partial y^{\mu_1}}\otimes\cdots\otimes\frac{\partial}{\partial y^{\mu_n}}\\
&=
D^{\mu_1\cdots\mu_n}
\frac{\partial}{\partial y^{\mu_1}}\otimes\cdots\otimes\frac{\partial}{\partial y^{\mu_n}},
\end{split}
\end{equation*}
of which 
\begin{equation}\label{eq-3-11}
D^{\mu_1\cdots\mu_n}
=
\underbrace{\int_0^t\cdots\int_0^t}_{n-1\ \text{tuples}} 
\left\langle \left\langle
\mathcal{L}^{\mu_1}\otimes\cdots\otimes\mathcal{L}^{\mu_n}
\right\rangle\right\rangle.
\end{equation}

For $n=1$, ${D}^{(1)} $ is a vector, expressed as 
\begin{equation*}
{D}^{(1)} 
=\frac{\partial \langle S^{\mu} \rangle}{\partial t}\frac{\partial}{\partial y^{\mu}}
=\langle\mathcal{L}^{\mu}\rangle\frac{\partial}{\partial y^{\mu}}
=D^{\mu}\frac{\partial}{\partial y^{\mu}},
\end{equation*}
of which $D^{\mu}=\langle\mathcal{L}^{\mu}(\dot{\gamma}(t))\rangle$ is the time rate of change of the averaged fibre length $\pi^{-1}(q)$, i.e., an averaged tangent vector to the stochastic fibre $\pi^{-1}(q)$, usually known as the convection of stochastic systems. 

For $n=2$, ${D}^{(2)} $ is a second-order tensor given by
\begin{equation*}
\begin{split}
{D}^{(2)} 
&=
\frac{1}{2!}\frac{\partial \langle\langle S^{\mu}S^{\nu} \rangle\rangle}{\partial t}\frac{\partial}{\partial y^{\mu}}\otimes\frac{\partial}{\partial y^{\nu}}\\
&=
\int_0^t  
\left\langle \left\langle
	\mathcal{L}^{\mu}(\dot{\gamma}(t))\mathcal{L}^{\nu}(\dot{\gamma}(\tau)) 
\right\rangle\right\rangle\mathrm{d}\tau
\frac{\partial}{\partial y^{\mu}}\otimes\frac{\partial}{\partial y^{\nu}}\\
&=
D^{\mu\nu}
\frac{\partial}{\partial y^{\mu}}\otimes\frac{\partial}{\partial y^{\nu}}, 
\end{split}
\end{equation*}
of which $D^{\mu\nu}=\int_0^t  
\left\langle \left\langle
	\mathcal{L}^{\mu}(\dot{\gamma}(t))\mathcal{L}^{\nu}(\dot{\gamma}(\tau)) 
\right\rangle\right\rangle\mathrm{d}\tau$. The second-order tensor $D^{\mu\nu}$ is due to the correlations of vectors and leads to diffusion in physics. It should be mentioned that 
\begin{equation*}
{D}^{(2)} 
=
D^{\mu\nu}
\frac{\partial}{\partial y^{\mu}}\otimes\frac{\partial}{\partial y^{\nu}}
=
D^{\mu\nu}
\left(
\frac{\partial^2}{\partial y^{\mu}\partial y^{\nu}}
+\frac{\partial}{\partial y^{\mu}}_{\frac{\partial}{\partial y^{\nu}}}
\right),
\end{equation*}
and  
\begin{equation*}
\frac{\partial}{\partial y^{\mu}}_{\frac{\partial}{\partial y^{\nu}}}
= A^{\sigma}_{\mu\nu}\frac{\partial}{\partial y^{\sigma}}, 
\end{equation*}
implying that the curvature $A^{\sigma}_{\mu\nu}$ can lead to drift of vector bundles.

For the terms of order $n\ge 3$, it is a $n$th-order tensor, given by
\begin{equation*}
{D}^{(n)} =
\frac{(-1)^n}{n!}D^{\mu_1\cdots\mu_n}\frac{\partial}{\partial y^{\mu_1}}\otimes\cdots\otimes\frac{\partial}{\partial y^{\mu_n}}.
\end{equation*}
Although the physical effects of $D^{(n)}$ for $n\ge 3$ on stochastic systems are not yet fully understood, we can infer from its expression that it is associated with more delicate statistical structures of stochastic fibre bundles. This suggests that $D^{(n)}$ captures intricate details of the system's dynamics and could play a significant role in describing complex stochastic phenomena.

With the help of the expressions for $D^{(n)}$ given above, the operator $\mathscr{L}$ is explicitly written as 
\begin{equation}\label{eq-3-12}
\begin{split}
\mathscr{L}
=
&-D^{\mu}\frac{\partial}{\partial y^{\mu}}
+\frac{1}{2!}D^{\mu\nu}\frac{\partial}{\partial y^{\mu}}\otimes\frac{\partial}{\partial y^{\nu}}
+\cdots\\
&+
\frac{(-1)^n}{n!}D^{\mu_1\cdots\mu_n}\frac{\partial}{\partial y^{\mu_1}}\otimes\cdots\otimes\frac{\partial}{\partial y^{\mu_n}}\cdots.
\end{split}
\end{equation}

\section{Geometric Structure of Stochastic Vector Bundles}\label{sec-4}
%
%
%

%
%
%
\subsection{\texorpdfstring{$\mathscr{L}$}{}-induced Jet Bundle on \texorpdfstring{$\mathbb{P}$}{}}\label{sec-4-1}
%
%
%


%
%
%
%
%
%

The operator $\mathscr{L}$ acts as a mapping that takes the function $P$ and produces an infinite order prolongation of $P$, given as follows \cite{saunders_1989}:
\begin{equation*}
\mathscr{L}: P \mapsto
\left\{
\frac{\partial}{\partial y^{\mu_1}}P, 
\frac{\partial}{\partial y^{\mu_1}}\otimes \frac{\partial}{\partial y^{\mu_2}}P, 
\cdots,
\frac{\partial}{\partial y^{\mu_1}}
\otimes\cdots\otimes 
\frac{\partial}{\partial y^{\mu_{\infty}}}
P
\right\}.
\end{equation*}
Let $j_{y}^{\infty}: P\mapsto j_y^{\infty}P$ be an infinite order prolongation of $P$, we can define an infinite order jet bundle 
$J_E^{\infty}\mathbb{P}=\{j_y^{\infty}P| y\in E, \ P\in \mathbb{P} \} $ with the canonical projections $\pi_{E,0}^{\infty}: J_E^{\infty}\mathbb{P}\to E$. In the terminology of jet bundle theory, it possesses the local coordinates $(y, p, p^{\mu_1}, p^{\mu_1\mu_2}, \cdots p^{\mu_1\cdots\mu_{\infty}}
)
$ which serve as a set of mappings:
\begin{equation*}
(y, p, p_{\mu_1}, p_{\mu_1\mu_2}, \cdots, p_{\mu_1\cdots\mu_{\infty}}
): 
j_y^{\infty}P\mapsto
\begin{dcases}
y(j_y^{\infty}P)=y\\
p(j_y^{\infty}P)=P(y),\\
p_{\mu_1}(j_y^{\infty}P)=\frac{\partial}{\partial y^{\mu_1}}P(y),\\
\cdots,\\
p_{\mu_1\cdots\mu_{\infty}}(j_y^{\infty}P)=\frac{\partial}{\partial y^{\mu_1}}\otimes\cdots\otimes\frac{\partial}{\partial y^{\mu_{\infty}}}
P(y). 
\end{dcases}
\end{equation*}

%
%
%
%
%
%

%

Let $X\in T(J^{\infty}_EP)$ be a tangent vector to $J^{\infty}_E\mathbb{P}$. With the coordinates $(y, p, p_{\mu_1},\cdots,p_{\mu_{\infty}})$, the tangent vector $X$ is expressed as
\begin{equation*}
X=
\dot{y}\cdot\frac{\partial}{\partial y}
+\sum_{|I|=0}^{\infty}\dot{p}_I \cdot\frac{\partial}{\partial {p}_{I} },
\end{equation*}
where $I=\mu_1\cdots\mu_n$ is a multi-index and with the length $|I|=n$ in this case. Dual to the tangent space of the jet bundle is the cotangent space $T^{*}J_E^{|I|}\mathbb{P} (0 \le |I|< \infty)$, on which a 1-form $\Theta \in T^{*}J_E^{|I|}\mathbb{P}$ is defined by
\begin{equation*}
\Theta=
\mathrm{d}p_I-
\Gamma(y, p_I)\mathrm{d}y, \ 0\le |I| < \infty, 
\end{equation*}
where $\Gamma(y, p_I)$ represents the local connection between $\mathrm{d}p_I$ and $\mathrm{d}y$. When  
\begin{equation*}
\Theta(X)=
\dot{p}_I-\Gamma(y, p_I)\dot{y} \equiv 0, \ 0\le |I| < \infty,
\end{equation*}
we found $X_H$, the horizontal part of $X$. It is
\begin{equation*}
X_H=
\dot{y}\cdot
\left[
\frac{\partial}{\partial y}+
\sum_{|I|=0}^{\infty}
\Gamma(y, p_I)\frac{\partial}{\partial {p}_{I}}
\right].
\end{equation*}
Correspondingly, the vertical part $X_V$ of $X$ is
\begin{equation*}
X_V=\sum_{|I|=0}^{\infty}
\left[
\dot{p}_I-\Gamma(y, p_I)\dot{y}
\right]
\frac{\partial}{\partial p_I}.
\end{equation*}
With $\{X_H\}$ and $\{X_V\}$, the tangent space $T(J_E^{\infty}\mathbb{P})$ is decomposed as 
\begin{equation*}
T(J_E^{\infty}\mathbb{P})=H(TJ_E^{\infty}\mathbb{P})\oplus V(TJ_E^{\infty}\mathbb{P}),
\end{equation*}
where  $H(TJ_E^{\infty}\mathbb{P})$ and $V(TJ_E^{\infty}\mathbb{P})$ are, respectively, the horizontal and vertical subspaces of  $T(J_E^{\infty}\mathbb{P})$. This decomposition is fundamental in analysing stochastic systems on jet bundles, as it separates the dynamics into those that do not affect the system (horizontal) and those that affect the fibre (vertical).

As a usual way, assuming that $\Gamma(y, p_I)$ is linear with respect to $p_I$, i.e, $\Gamma(y, p_I)=\Gamma(y)p_I$, we had that  
\begin{equation*}
\Theta=
\mathrm{d}p_I-
[\Gamma(y)\mathrm{d}y] p_I, \ 0\le |I| < \infty.
\end{equation*}
$\Theta$ determines a series of curves on $E$. Let $u=\dot{y}\cdot\partial_y$ be the tangent vector to these curves, then $\Theta(u)=0$ results in the parallel transport equations:
\begin{equation*}
\left\{
\frac{\partial }{\partial t}
- [\Gamma(y)\mathrm{d}y](u)
\right\}p_I
=0, \ 0\le |I| < \infty,
\end{equation*}
where we used $\mathrm{d}p_{I}(u)=\dot{y}\cdot\partial p_I/\partial y=\mathrm{d}p_I(y)/\mathrm{d}t$. At the same time, using the jet map
\begin{equation*}
p_I(j_{y}^{\infty}P)=\frac{\partial}{\partial y^{\mu_1}} \otimes \cdots \otimes\frac{\partial}{\partial y^{\mu_{|I|}}}P(y),
\ 0\le |I| < \infty,
\end{equation*}
which associates $p_I$ with $P$, we derived the geodesic equations for $J_E^{\infty}\mathbb{P}$:
\begin{equation}\label{eq-4-1}
\left\{\frac{\partial}{\partial t}-[\Gamma(y)\mathrm{d}y](u)\right\}
\left(
\frac{\partial}{\partial y^{\mu_1}}\otimes\cdots\otimes\frac{\partial}{\partial y^{\mu_{|I|}}}\right)P
=0, \ 0\le |I| < \infty.
\end{equation}

When $|I|=0$, Eq.~\eqref{eq-4-1} reduces to 
\begin{equation}\label{eq-4-2}
\left[\frac{\partial}{\partial t}-\Gamma(y)(\dot{y})\right]
P
=\mathrm{d}P(\dot{y})-\Gamma(y)(\dot{y})P(y)
=0,
\end{equation}
where the identities $\mathrm{d}P(\dot{y})=\mathrm{d}P(y)/\mathrm{d}t$ and $\mathrm{d}y(u)=\dot{y}$ are applied in the derivation. Eq.~\eqref{eq-4-2} can also be regarded as the geodesic equation for $y(t) \in E$  with a given probability $P$, which gives the path of the shortest distance between two points in $E$. Furthermore, since $\dot{P}$ also satisfies Eq.~\eqref{eq-3-10-a}, by comparing the above geodesic equation and Eq.~\eqref{eq-3-10-a}, we had that  
\begin{equation*}
\mathscr{L}(\dot{\gamma}(t))
=\Gamma(y)(\dot{y}(t)).
\end{equation*}
Since the projections $\pi$ induces a push-forward $\pi_{*}$ acting on the tangent vector $\dot{y}$ yields the tangent vector $\dot{\gamma}$, i.e., $\pi_{*}\dot{y}=\dot{\gamma}$, we established the relationship of $\mathscr{L}$ and $\Gamma$ as
\begin{equation*}
\mathscr{L}(\pi_{*}\dot{y})
=
\pi^{*}\mathscr{L}(\dot{y})
=\Gamma(y)(\dot{y}),
\end{equation*}
or simply,
\begin{equation*}
\Gamma(y)=\pi^{*}\mathscr{L},
\end{equation*}
which verifies that $\mathscr{L}$ indeed yields a well-defined jet bundle and acts as the connection of the jet bundle.

In Section \ref{sec-3}, we presented a formulation of the kinetic equation in a physical way in vector bundles, while in this part of the paper, we derived it as a geodesic equation of jet bundle (Eq.~\eqref{eq-4-2}) from the viewpoint of differential geometry.  The significance of the geometric formulation lies in the fact that it provides a new angle to view kinetic equations.  Most importantly, the above formulations prove that the kinetic equation Eq.~\eqref{eq-3-10-a} represents the geodesic equation in the probability space, which is generic and topologically invariant. This observation is crucial for investigating stochastic problems on general manifolds.

\subsection{Structure of Stochastic Vector Bundles}\label{sec-4-2}

The preceding formulation reveals that the relationship between configuration space $M$ and the jet bundle $J_E^{\infty}\mathbb{P}$ is established through a sequence of jet bundles. This sequence provides a comprehensive perspective on the geometric structure of stochastic processes, elucidating their relationship through the fibre bundle theory. The sequence of surjective projections are: 
\begin{equation*}
J^{\infty}_E\mathbb{P} \xrightarrow{\pi_{E}^{\infty}}\mathbb{P}\xrightarrow{\pi_{E}}E\xrightarrow{\pi}M,
\end{equation*}
which carries a physical interpretation as follows:
\begin{itemize}
\item[1.] The manifold $M$, which is the base manifold of the vector bundle $E$ related by the projection $\pi: E\to M$, encapsulates the ensemble of stochastic processes $\gamma(t)$.

\item[2.] The vector bundle $E$ constitutes phase space for $\gamma(t)$; each fiber $\pi^{-1}(\gamma(t)) \in E$ is a collection of the states of the stochastic process $\gamma$ at time $t$.

\item[3.] The jet bundle $\pi^{\infty}_{E}: J_E^{\infty}\mathbb{P}\to \mathbb{P}$  encodes the statistical dynamics of the stochastic processes, governed by the geodesic equations $\dot{P}=\mathscr{L}P$.
\end{itemize}

At the same time, the kinetic equation has a formal solution as
\begin{equation*}
P(t)
=g(t) P(0),
\end{equation*}
where $g\in \mathbb{G}$, and $\mathbb{G}$ is a group defined as
\begin{equation*}
\mathbb{G}
 =\left\{
 \mathrm{e}^{\int\pi^{*}\mathscr{L}(y(t))\mathrm{d}t}\bigg{|}\forall y\in E 
 \right\},
\end{equation*}
which determines an equivalent relation. Obviously, in the case of a discrete stochastic system situated on an $n$-dimensional vector space, the state transition probabilities can indeed be represented by a matrix $P=\{P_{ij}\}_{n\times n}$, which belongs to the general linear group $GL(n, \mathbb{R})$. Consequently, the projection map $\pi_E: \mathbb{P}\to E$ defines a principal bundle, where $\mathbb{G}$ acts as the structure group and the typical fibre. This construction allows for the geometric representation of the system's probabilistic transitions within the framework of principal bundle theory.  A detailed commutative diagram regarding the relation mentioned above is shown in Fig.~\ref{fig: Fig2}, which depicts the geometric structure of the stochastic vector bundle:  
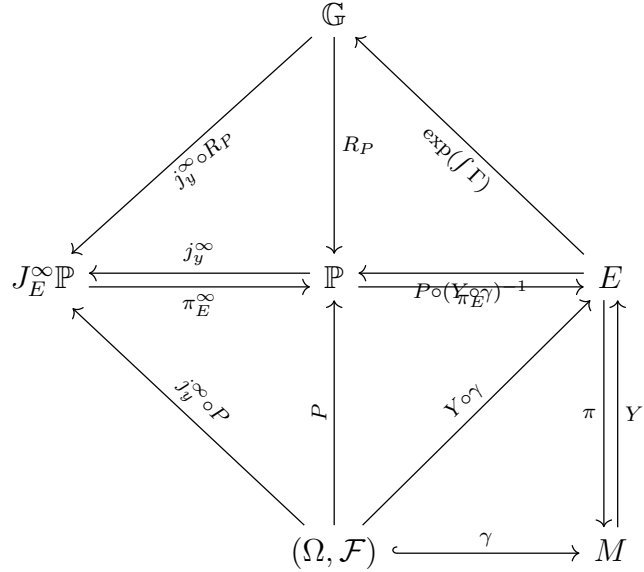
\begin{figure}[h]
\centering
\begin{tikzcd}[ column sep=6em, row sep=7em]
&\mathbb{G} 
	\arrow[sloped,swap]{ld}{j_y^{\infty}\circ R_P}
	\arrow{d}{R_P}
&\\
J_{E}^{\infty}\mathbb{P}
	\arrow[yshift=-0.5ex,swap]{r}{\pi^{\infty}_E}
&{\mathbb{P}}
	\arrow[yshift=0.5ex,swap]{l}{j_y^{\infty}}
	\arrow[yshift=-0.5ex,swap]{r}{\pi_E}
&E
	\arrow[yshift=0.5ex,sloped,swap]{l}{P\circ (Y \circ \gamma)^{-1}}
	\arrow[xshift=-0.5ex,swap]{d}{\pi}
	\arrow[sloped,swap]{lu}{\mathrm{exp}(\int \Gamma)}
	\\
&(\Omega, \mathcal{F})
	\arrow[hook]{r}{\gamma}
	\arrow[sloped]{ur}{Y\circ\gamma}
	\arrow[sloped]{u}{P}
	\arrow[sloped]{lu}{j_y^{\infty}\circ P}
&M
	\arrow[xshift=0.5ex,swap]{u}{Y}
\end{tikzcd}
\caption{\label{fig: Fig2} The commutative diagram for the jet bundle  $\pi_{E, 0}^{\infty}: J_{E}^{\infty}\mathbb{P}\to E$, $\pi_{E, 0}^{\infty}=\pi_E\circ\pi_E^{\infty}$. 
$\gamma$ is a stochastic processes defined on $(\Omega, \mathcal{F})$; $Y: M\to E$ is the section of $\pi: E\to M$; $P: (\Omega, \mathcal{F})\to \mathbb{P}$ is the probability measure; $j_y^{\infty}$ is the jet of $P$ that $j_{y}^{\infty}: P\mapsto j_{y}^{\infty}P \in J_{E}^{\infty}\mathbb{P}$; $R_P:  \mathbb{G}\to  \mathbb{P}$ is an inclusion; the mapping $\exp(\int \Gamma): E\to \mathbb{G}$ constitutes a group $\mathbb{G}$ acting on $\mathbb{P}$, and $\Gamma=\pi^{*}\mathscr{L}$.
}
 \end{figure}

\section{Discussions}\label{sec-5}
\subsection{Non-Markovianity}\label{sec-5-1}
Non-Markovian behaviours in stochastic processes are essential because they reflect that many systems have memory, meaning their future states can depend on more than their current state. In contrast, a Markov process is memoryless, where the future states only rely on the current state and not on the history of how the process arrived at that state. This memory effect can be critical for understanding and predicting the behaviour of complex systems in various fields, including physics, finance, biology, and engineering. 

For divergence-free systems, since that 
\begin{equation*}
\frac{\partial \dot{Z}^{\mu}}{\partial z^{\mu}}=0.
\end{equation*}
The evolution equation for $\chi$ can be written as 
\begin{equation*}
\frac{\partial \chi}{\partial t}=-\frac{\partial }{\partial z^{\mu}}\dot{Z}^{\mu} \chi,
\end{equation*}
and following the same procedure in deriving Eq.~\eqref{eq-3-10}, 
we derived a kinetic equation that is less geometric in form but presented in a way that is more familiar to researchers working in mechanics or physics. This form allows for an easier interpretation and application of the equation within the existing frameworks of these disciplines. It is:
\begin{equation}\label{eq-5-1}
\frac{\partial f(y, t|y_0, 0)}{\partial t}=\sum_{n=1}^{\infty}\frac{ (-1)^n}{n!} \partial_{y}^n {D}^{(n)}f(y, t|y_0, 0),
\end{equation}
where
\begin{equation*}
\partial_{y}^n=\underbrace{\partial_y\otimes \cdots \otimes \partial_y}_{n \ tuples},
\end{equation*}
and 
\begin{equation*}
{D}^{(n)}
=D^{\mu_1\cdots\mu_n}.
\end{equation*}
Eq.~\eqref{eq-5-1} is usually termed as the conservative form of Eq.~\eqref{eq-3-10-b}. Noticed that 
\begin{equation*}
\frac{\partial f(y, t)}{\partial t}=\int  f(y_0,0)\frac{\partial f(y, t|y_0, 0)}{\partial t} \mathrm{d}y_0,
\end{equation*}
with the similar method has been adopted in \cite{ZHONG-2022-KINETICEQAUTION}, one finds that 
\begin{equation}\label{eq-5-2}
\begin{split}
\frac{\partial f(y, t)}{\partial t} 
=
-\frac{\partial}{\partial{y^{\mu}}} {V}^{\mu} f(y, t) 
+
\frac{\partial}{\partial{y^{\mu}}} \int_0^t \int \mathcal{K}^{\mu\nu}(y, y_0, t-s)\frac{\partial}{\partial{y^{\nu}_0}}f(y_0, s) \mathrm{d}s \mathrm{d}y_0
+\cdots,
\end{split}
\end{equation}
where 
\begin{equation*}
{V}^{\mu}=\frac{1}{f(y, t)}\int  D^{\mu} f(y, t|y_0, 0) f(y_0, 0) \mathrm{d}y_0
\end{equation*}
is the drift velocity, and 
\begin{equation*}
\mathcal{K}^{\mu\nu}(y, y_0, t-s)=\frac{1}{2!}D^{\mu\nu}f(y, t|y_0, s)
\end{equation*}
is the historical kernel. If truncated at n=2, Eq.~\eqref{eq-5-2} has the same form as that derived in \citep{zwanzig2001nonequilibrium}, which is regarded as a typical modification of the classical FPE to non-Markovian processes.

The vectors $V^{\mu}$ and the historical integration kernel $\mathcal{K}^{\mu\nu}(y, y_0, t-s)$ are critical elements in capturing the dynamics of stochastic systems with memory: 

First, the vector $V^{\mu}$ represents the conditional average velocity derived from all possible state transition trajectories that start at $y_0$ and end at $y$. This conditional average reflects the historical influences on the system's drift between different state distributions, encoding information from the entire ensemble of state transition trajectories.

Second, the historical integration kernel $\mathcal{K}^{\mu\nu}(y, y_0, t-s)$ is defined as $\frac{1}{2}D^{\mu\nu}f$. It plays a critical role in incorporating the historical effects within stochastic systems through two primary mechanisms:
(1) $D^{\mu\nu}(y, t|y_0, s)$, which is influenced by the cumulants of the integrated state transition paths. This term captures the statistical correlations and fluctuations in the system's dynamics over time, and (2) $f(y, t|y_0, s)$, which modulates $D^{\mu\nu}(y, t|y_0, s)$ through a conditional average. This conditional average accounts for the probabilities of different state transitions and their associated velocities, effectively weighting the influence of various paths on the system's overall behaviour.

These mechanisms are inherent in $\mathcal{K}^{\mu\nu}(y, y_0, t-s)$ and can either amplify or dampen the intrinsic irregularities and fluctuations observed in stochastic processes. By incorporating both the state transition intensity (i.e., $f(y,t|y_0, 0)$) and the historical correlations (i.e., $D^{\mu\nu}(y,t|y_0, 0)$), the kernel provides a more nuanced understanding of the system's evolution, allowing for a more accurate description of the complex interplay between historical influences and current dynamics in stochastic systems.

%
%
%
%
%
%

%
%
%
\subsection{Markov Approximation}\label{sec-5-2}

In the context of this paper, the Markov approximation is invoked when the correlation time scale $T_L$ is of small magnitude, suggesting that the system's memory is short and the future depends primarily on the present state. Under this consideration, it is valid to use the following simplification:
\begin{equation*}
S=\int_0^t \mathrm{d}\tau \mathcal{L}(\dot{\gamma}(\tau))
\approx \mathcal{L}(\dot{\gamma}(t))T_L.
\end{equation*}
The consideration assumes that the average effect of the operator $\mathcal{L}$ over the time interval $T_L$ is equivalent to its instantaneous value at time $t$ multiplied by the correlation time $T_L$. 

Based on this consideration, the paper discusses the Markov approximations for flat vector spaces and vector bundles. For flat vector spaces, the familiar classical FPE can be derived. In the context of vector bundles, the Markov approximation helps to simplify the kinetic equation, making it more tractable for analysis and computation while still capturing the essential dynamics of the system.

\subsubsection{Markov approximation for Euclidean space}\label{sec-5-2-1}

For flat space, $\omega_{\gamma}=0$ as discussed in subsection \ref{sec-3-3}, the connection 1-form $\omega=\mathrm{d}+\omega_{\gamma}$ is reduced to 
\begin{equation*}
\omega=\mathrm{d},
\end{equation*}
so that 
\begin{equation*}
\begin{split}
\mathcal{L}(\dot{\gamma}(\tau))
&=Y\otimes \omega(\dot{\gamma}(\tau)) \\ \nonumber
&= \mathrm{d}Y(\dot{\gamma}(\tau))+
\underbrace{(P_{\gamma}^{-t}\mathrm{d}P_{\gamma}^{t}(\dot{\gamma}(\tau))
+\mathrm{Ad}_{P_{\gamma}^{-t}}A_{\gamma}(\dot{\gamma}(\tau)))}_{=\omega_{\gamma}=0 \text{\ for flat space}}Y\\ \nonumber
&=\mathrm{d}Y(\dot{\gamma}(\tau))=
\dot{Y}(\tau|y_0,0)
 \end{split}
\end{equation*}
and 
\begin{equation*}
\frac{\partial}{\partial y^{\mu_1}}\otimes\cdots\otimes \frac{\partial}{\partial y^{\mu_n}}
=\frac{\partial^n}{\partial y^{\mu_1}\cdots\partial y^{\mu_n}}.
\end{equation*}
At the same time, for Markovian processes, the correlation time scale $T_L$ is small, implying 
\begin{equation*}
S=\int_0^t \mathrm{d}\tau \mathcal{L}(\dot{\gamma}(\tau))
\approx \mathcal{L}(\dot{\gamma}(t))T_L
= \dot{Y} T_L,
\end{equation*}   
and
\begin{equation*}
\begin{split}
D^{\mu_1\cdots\mu_n}
=
\frac{1}{n!}\frac{\partial}{\partial t}\frac{\langle\langle S^n\rangle\rangle}{n!}
\approx
\frac{1}{n!}\frac{\langle\langle S^n(t)\rangle\rangle}{T_L}.
\end{split}
\end{equation*}
For example, for $n=1$, 
\begin{equation*}
D^{\mu}
=
\frac{\partial \langle S\rangle}{\partial t}
\approx \frac{\langle S(t)\rangle}{T_L}=\langle \dot{Y}(t)\rangle,
\end{equation*}
and for $n=2$, 
\begin{equation*}
D^{\mu\nu}
\approx
\frac{1}{2!}\frac{\partial}{\partial t}\frac{\langle\langle S^{\mu}S^{\nu}\rangle\rangle}{2!}
=
\frac{1}{2!}\frac{\langle \delta S^{\mu} \delta S^{\nu}\rangle -\langle  \delta S^{\mu}\rangle \langle \delta S^{\nu} \rangle}{T_L},
\end{equation*}
where $\delta S = S-\langle S\rangle$. Noticed that for Markovian processes, $\langle\delta S^{\mu}\delta S^{\nu}\rangle \sim O(T_L)$, but $\langle \delta S^{\mu}\rangle\langle \delta S^{\nu}\rangle \sim O(T_L^2)$, resulting in
\begin{equation*}
\begin{split}
D^{\mu\nu}
\approx
\frac{1}{2}\langle \dot{Y}^{\nu}(t) \dot{Y}^{\mu}(t-T_L)\rangle T_L.
\end{split}
\end{equation*}
For stable stochastic processes, it is 
\begin{equation*}
\begin{split}
D^{\mu\nu}
\approx
\frac{1}{2}\langle \dot{Y}^{\mu}(0) \dot{Y}^{\nu}(T_L)\rangle T_L.
\end{split}
\end{equation*}
High-order terms ($n\ge 3$) can be derived similarly. When truncated at $n=2$, we obtained the classical FPE for $\mathbb{R}^n$:
\begin{equation}\label{eq-5-3}
\frac{\partial f}{\partial t}
+D^{\mu} \frac{\partial f}{\partial y^{\mu}} 
=D^{\mu\nu}\frac{\partial^2 f}{\partial y^{\mu}\partial y^{\nu}}.
\end{equation}

\subsubsection{Markov approximation for vector bundles}\label{sec-5-2-2}
For the cases of a Markovian stochastic process on curved space
\begin{equation*}
\begin{split}
\frac{\partial}{\partial y^{\mu_1}}\otimes\cdots\otimes \frac{\partial}{\partial y^{\mu_n}}
&=
\frac{\partial^n}{\partial y^{\mu_1}\cdots\partial y^{\mu_n}}\\
&+\frac{\partial}{\partial y^{\mu_1}}\left(\frac{\partial}{\partial y^{\mu_2}}\otimes\cdots\otimes \frac{\partial}{\partial y^{\mu_n}}
\right)\\
&+\cdots,
\end{split}
\end{equation*}
and
\begin{equation*}
S
=\int_0^t \mathcal{L}(\dot{\gamma}(\tau))\mathrm{d}\tau 
\approx \mathcal{L}(\dot{\gamma}(t))T_L.
\end{equation*} 

For $n=1$, 
\begin{equation*}
\begin{split}
D^{\mu}
=
\frac{\partial}{\partial t}\langle S^{\mu}\rangle
\approx
\langle\mathcal{L}^{\mu}(\dot{\gamma}(t)) \rangle,
\end{split}
\end{equation*}
and for $n=2$, 
\begin{equation*}
\begin{split}
D^{\mu\nu}
\approx
\frac{1}{2}\langle \mathcal{L}^{\mu}(\dot{\gamma}(t)) \mathcal{L}^{\nu}(\dot{\gamma}(t-T_L))\rangle T_L.
\end{split}
\end{equation*}

Then, the FPE for vector bundle $E$ is 
\begin{equation}\label{eq-5-4}
\begin{split}
\frac{\partial f}{\partial t}
+D^{\mu} \frac{\partial f}{\partial y^{\mu}} 
-D^{\mu\nu} A^{\sigma}_{\mu\nu}\frac{\partial f}{\partial y^{\sigma}}
=D^{\mu\nu}\frac{\partial^2 f}{\partial y^{\mu}\partial y^{\nu}}. 
\end{split}
\end{equation}

Eq.~\eqref{eq-5-4} reduces to Eq.~\eqref{eq-5-3} for $\mathbb{R}^n$, of which $\mathcal{L}(\dot{\gamma}(t))=\dot{Y}(t)$ and $A^{\sigma}_{\mu\nu}=0$; moreover, Eq.~\eqref{eq-5-4}, specifically, the third term on the right-hand side of the equation, shows that randomness in vector fibres ($D^{\mu\nu} $) together with curvatures of the vector bundles ($A^{\sigma}_{\mu\nu}$) leads to an averaged drift effect.  

The Markovian approximation can also be derived using Eq. (5.2). It is clear that $y_0\to y$ and $\partial/\partial y_0\to \partial/\partial y$ in a Markovian process, suggesting that the system's behaviour is independent of its history, which is a characteristic of Markov processes. Moreover, the integration kernel decreases rapidly, and we only need to integrate the kernel with respect to time \cite{zwanzig2001nonequilibrium}. Therefore, 
\begin{equation*}
\begin{split}
\int_0^t  \int \mathcal{K}^{\mu\nu}(y, y_0, t-s)\frac{\partial}{\partial{y^{\nu}_0}}f(y_0, s) \mathrm{d}s\mathrm{d}y_0
&\approx
\int_0^t   \mathcal{K}^{\mu\nu}(y, y, t-s)\frac{\partial}{\partial{y^{\nu}}}f(y, s) \mathrm{d}s  \\
&=
\frac{\partial}{\partial{y^{\nu}}} \int_0^t  \mathcal{K}^{\mu\nu}(y, y, t-s)f(y, s) \mathrm{d}s \\
&=\frac{1}{2!}\frac{\partial}{\partial{y^{\nu}}}  \int_0^t D^{\mu\nu}\delta(t-s)f(y,s)\mathrm{d}s\\
&=\frac{1}{2!}\frac{\partial}{\partial{y^{\nu}}}  D^{\mu\nu}f(y,t)
\end{split}
\end{equation*}
In the derivation above, we used the identity $f(y,t|y,s)=\delta(t-s)$, of which $\delta$ is the Dirac Delta function.

\subsection{Macroscopic Conservation Equations}\label{sec-5-3}

The kinetic approach is extensively utilised in exploring diverse stochastic phenomena, with notable applications in fluid dynamics, as detailed in \cite{SBPOPE-TURBULENTFLOWS}. Beyond its use in analysing the statistical characteristics of turbulence, the kinetic equation is further applied to deduce macroscopic conservation equations for complex multiphase flows, as demonstrated in \cite{zhang1994averaged,zhang1994ensemble,zhang1997momentum, ZHONG-2022-KINETICEQAUTION}.

Let $\psi$ be a microscopic dynamical variable, for instance, the density of fluid particles or the velocities of fluid particles. Define a conditional ensemble average of $\psi$ as
\begin{equation*}
 f(y,t) \langle \psi \rangle=\int  \psi(y,t|y_0,0) f(y,t|y_0,0) \mathrm{d}y_0, 
\end{equation*}
and differentiate it with respect to time $t$, and with the help of the kinetic equation, we find that 

\begin{eqnarray} \label{eq-5-5}
\frac{\partial f(y ,t) \langle \psi  \rangle }{\partial t} =  
\mathscr{L}(f(y,t)\langle\psi \rangle)  
 +  
 f(y ,t) \left\langle   \dot{\psi}  \right\rangle 
\end{eqnarray}
where
\begin{eqnarray*}
f(y ,t) \left\langle   \dot{\psi}  \right\rangle   
= \int  
\left(\frac{\partial   \psi    }{\partial t} 
-\mathscr{L}   \psi    \right)  f(y,t|y_0,0) \mathrm{d} y_0
\end{eqnarray*}

Furthermore, for fluid flows, $y=(r, v)$, where $r$ is the position vector and $v$ is the velocity vector, we define that 
\begin{equation*}
\overline{\psi}=\int  f(r,v,t)\langle \psi(r,v,t)\rangle \mathrm{d}v
\end{equation*}
Integrating Eq.~\eqref{eq-5-5} with respect to $v$ and noticing that $f(r,v,t)\to 0$ when $v\to \pm \infty $, we had that   
\begin{equation} \label{eq-5-6}
\frac{\partial \overline{\psi}}{\partial t}
+
\overline{\mathscr{L}_r \langle \psi\rangle}
=\overline{\left\langle   \dot{\psi}  \right\rangle}
\end{equation}
where $\mathscr{L}_r$ is the time evolution operator with respect to space coordinate $r$ having the same form as $\mathscr{L}$ but with $y^{\mu}$ replaced by $r^{\mu}$.  For $\psi=\rho$ and $\psi=\rho v$, we obtained mass and momentum conservation equations for fluid flows, respectively. Assuming that the motion of fluid particles is a Markovian process and truncating at the second order, we obtain the famous Navier-Stokes equations. It can be seen that the second term on the left-hand-side of Eq.~\eqref{eq-5-6} represents the non-linear convection of flow, while the term on the right-hand-side of it represents internal and external actions on fluids.

The kinetic approach is also applicable to multiphase flow systems. Pioneering works by Reeks \citep{Reeks1991Kinetic} and others \cite{SWAILES199738, Hyland1999PDF, ZHONG-2022-KINETICEQAUTION, zhang1994averaged} have utilised kinetic theory to address specific problems in multiphase flows. These studies often involve the kinetic equation (such as the Fokker-Planck equation, the Boltzmann equation, and their variants) adapted for the stochastic system by ascribing physical meaning to the state transition path $Y$. The kinetic equation proposed in this paper is ready to analyse multiphase flows.

This begins with the definition of the local density function. If we let $Y_k=(R_k, V_k)$ where $R_k$ and $V_k$ are, respectively, the position vector of $k^{th}$-phase particles and the velocity vector of kth-phase particles, then
\begin{equation*}
\chi_k(Y_k, y)=\chi(|Y_k(t|y_{0}, 0)-y|)
=
\begin{cases}
\infty, &Y_k(t|y_{k0}, 0)=y,\\
0, &otherwise,
\end{cases}
\end{equation*} 
determines if phase point $y=(r, v)$ is occupied by the particles of $k^{th}$-phase. By this definition, we find that 
\begin{equation*}
f(t, r_k, v_k|0, r_{k0}, v_{k0})=\langle \chi_k(|R_k- r|)\chi_k (|V_k-v|) \rangle
\end{equation*}
is the conditional PDF for multiphase systems. Consequently, following the same procedure for general systems described above, we can derive a kinetic equation for multiphase systems.  

It deserves to be mentioned that the PDF formulations of turbulent single-phase flows and multiphase flows also provide an approach for closing the correlation term $\overline{v_k \psi_k}$ and source term $\overline{\left\langle   \dot{\psi}_k  \right\rangle}$. Details can be found in  \citep{Reeks1991Kinetic, reeks1992continuume, Hyland1999PDF, zaichik2004probability, reeks_2005, ZHONG-2022-KINETICEQAUTION,zhang1994averaged}.

\subsection{Gauge Transformation}\label{sec-5-4}
We heavily rely on the concepts developed in differential geometry to derive the kinetic equation. These geometric operations have profound physical meanings. The gauge transformation is one of the most significant ones, given by
$$
\omega_{\gamma}=P_{\gamma}^{-t}\mathrm{d}P_{\gamma}^{t}+\mathrm{Ad}_{P_{\gamma}^{-t}} A_{\gamma},
$$
of which the local connection 1-form $A_{\gamma}$ is physically understood as the gauge potential in particle physics, the critical concept in the well-known  Yang-Mills theory. Mathematically, $A_{\gamma}$, in this paper, leads to an infinitesimal translation of geometric objects by 
\begin{equation*}
P_{\gamma}^t=\overleftarrow{T} \mathrm{exp}\left(-\int A_{\gamma}\mathrm{d}\tau\right),
\end{equation*}
which moves a vector in a parallel manner. In our study, gauge transformation enables us to compare two vectors in different fibres so that the covariant derivative of a stochastic vector defined on a fibre bundle can be derived.

Related to the parallel translation $P_{\gamma}^t$, the other two important concepts used in this study are push-forward and pull-back. Push-forward typically refers to ``pushing" a geometric object, such as a vector, to another position or shape through a mapping or continuous transformation between manifolds. Conversely, pull-back is ``pulling back" a geometric object to the original reference configuration. In our study, push-forward $\psi_{*}=P_{\gamma}^t$ and correstponding pull-back $\psi^{*}$ are used to move the Liouville operator $\mathcal{L}$, a vector-valued 1-form, forward and backward, respectively. By these operations, we establish the relationship of $\mathcal{L}$ between the starting points and the ending points of system transitions, through which we can find the coordinate-independent expression for the time evolution operator $\mathscr{L}$ with historical effects accounting for.  

These differential geometry tools were also adopted in studies of fluid dynamics accompanied by intrinsic geometry problems. It has also been adopted in the study of Shapere and Wilczek \cite{Shapere_Wilczek_1989}, where the authors discussed the general kinematic framework for self-drive problems at low Reynolds numbers. The authors formulated the issue of self-propulsion as a gauge field problem on shape space. Shape space can be regarded as a manifold, where different shapes of an object are seen as points on this manifold. The concept of fibre bundles describes changes in the object's shape, where the base space is a collection of shapes, and the fibres represent all possible rigid body motions. They employed concepts from gauge field theory to describe the net translation and rotation of an object due to shape changes. In addition, in the context of the paper of Shapere and Wilczek \cite{Shapere_Wilczek_1989}, when considering infinitesimal changes in the shape of an object, the push-forward operation was used to describe how these changes affect the fluid flows, and the pull-back operations assist in understanding the impact of shape changes on fluid dynamics from a fluid dynamics perspective. They found that when the boundary of an object deforms, the push-forward and pull-back helped understand how the fluid moves under the new boundary conditions and how this movement impacts the net motion of the object.

Gauge transformation is an essential mathematical tool for handling general manifolds, which enables us to establish a general framework for complicated dynamical systems.

\subsection{Truncation of Infinite-order Kinetic Equation}\label{sec-5-5}

The kinetic equation formulated in this paper is of infinite order in its partial derivatives. A fundamental question arises concerning the number of terms that should be retained for practical applications. Determining where to truncate the infinite series of the kinetic equation has been an ongoing challenge yet to be resolved entirely.

The classical FPE represents a second-order approximation derived from the Kramer-Moyal expansion using the Pawula theorem \cite{risken1984fokker}. However, as indicated by \cite{risken1984fokker}, higher-order terms, specifically those with $n \geq 3$, can influence the PDF. Consequently, in certain scenarios, it is necessary to retain these higher-order terms to enhance the precision of predictions and to capture crucial phenomena that are associated with these terms. For example, the third-order derivative often characterises the skewness of the distribution. Incorporating third-order terms into the equation can reveal unique phenomena. The most important ones that could be deduced are:
\begin{enumerate}
\item  Anomalous diffusion: the third-order term can cause anomalous diffusion by long-tail distribution, resulting in subdiffusion or superdiffusion.
\item  Non-Gaussianity: the probability distribution will become non-Gaussian, even if initially Gaussian, due to the presence of the third-order term.
\item  Non-Markovian dynamics: the third-order term breaks the Markov property, leading to non-Markovian dynamics.
\end{enumerate}

Including third-order and higher terms significantly broadens the range of phenomena that can be described by the truncated kinetic equation, enabling it to model more complex stochastic systems. However, if third-order and higher terms are considered, it will also lead to significant difficulties, including computational complexity, numerical instabilities, parameter estimation challenges, etc. Therefore, theoretical answers to how many terms should be kept in applications are still open, waiting for mathematicians to give solid criteria to balance accuracy and feasibility.  

Nevertheless, Reeks \cite{Reeks1991Kinetic} and Swailes and Darbyshire \cite{SWAILES199738} suggested that, given the limited knowledge regarding the random forcing, keeping the terms for $n\le 2$ is advisable from a practical point of view. Fortunately, the high-order cumulants contain information on decreasing importance, unlike high-order moments that contain information about lower moments \cite{Sto-m-cgispinGardiner}, implying that the truncation of high-order terms may not lead to apparent errors. This point was also observed in our study reported in \cite{ZHONG-2022-KINETICEQAUTION}, where the Reynolds stresses of particles are calculated in turbulent flows, and satisfactory results were obtained.

\section{Conclusions}\label{sec-6}

The understanding of stochastic systems dramatically relies on kinetic equations. Throughout the years, a multitude of these equations have been created, incorporating different theories, methodologies, and mathematical techniques. However, when dealing with stochastic systems on a global scale, it becomes essential to explore stochastic systems on general manifolds rather than solely focusing on subspace embedding in Euclidean space. Our recent emphasis has been on developing a kinetic equation for stochastic vector bundles to broaden local stochastic analysis into global analysis. The significant results of our work include:

\begin{enumerate}
\item A stochastic vector bundle is a mathematical construct that combines a vector bundle with a probability space enriched by infinite-order jets. This structure provides a framework for addressing stochastic phenomena on manifolds.

\item The kinetic equation, when viewed as a geodesic equation on the jet bundle of the probability space, acquires properties of coordinate independence and geometric invariance. This characteristic is especially crucial for the analysis of general stochastic processes.

\item  The kinetic equation applies to non-Markovian processes, encompassing the classical FPE as a specific instance for Markovian processes within a flat Euclidean space. Its innate global geometric properties enable it to investigate stochastic phenomena from a global perspective.

\item Utilizing the infinite-order kinetic equation in real-world scenarios presents a formidable challenge. To date, our knowledge about this matter is restricted, and it is recommended from a practical standpoint that the terms of drift and diffusion should be maintained when applying the equation. Although this approximation is widely practised and has shown some level of precision, more rigorous mathematical examination and extensive numerical testing are crucial to confirm and improve this approximation. Furthermore, it might be beneficial to investigate under what conditions an infinite-order jet can be effectively represented by a finite-order jet, considering that the kinetic equation is an outcoming of an infinite-order jet.

\end{enumerate}

%
%
%



\section*{Acknowledgements}
This study was supported by the National Natural Science Foundation of China, Grant No. 91547204.
The authors have no conflict of interest or competing interests to declare that are relevant to the content of this article.

\section*{Data availability statement}
No new datasets were created or analysed in this study.

\bibliography{bib2019.bib}

\end{document}